\documentclass[twocolumn,english,american,aps,amsfonts,amsmath,prd,tightenlines,nofootinbib]{revtex4}
\usepackage[T1]{fontenc}
\usepackage[latin1]{inputenc}
\usepackage{amssymb}
\usepackage{amsmath}

\makeatletter

\usepackage{epsf}

\def\be{\begin{equation}}
\def\ee{\end{equation}}
\def\beq{\begin{equation}}
\def\eeq{\end{equation}}
\def\bea{\begin{eqnarray}}
\def\eea{\end{eqnarray}}
\def\bml{\begin{subequations}}
\def\blea{\bml\begin{eqnarray}}
\def\elea{\end{eqnarray}\end{subequations}}
\def\nn{\nonumber}

\usepackage{babel}
\makeatother
\begin{document}

\title{How many universes are in the multiverse?}

\selectlanguage{english}

\author{Andrei Linde}

\selectlanguage{american}

\email{alinde@stanford.edu}

\selectlanguage{english}

\author{Vitaly Vanchurin}

\selectlanguage{american}

\email{vanchurin@stanford.edu}

\selectlanguage{english}

\date{\today}

\affiliation{Department of Physics, Stanford University, Stanford, CA 94305, USA}

\selectlanguage{american}
\begin{abstract}
We argue that the total number of distinguishable locally Friedmann ``universes'' generated by eternal inflation is proportional to the exponent of the entropy of inflationary perturbations and is limited by $e^{e^{3 N}}$, where $N$ is the number of e-folds of slow-roll post-eternal inflation. For simplest models of chaotic inflation, $N$ is approximately equal to de Sitter entropy at the end of eternal inflation; it can be exponentially large. However, not all of these universes can be observed by a local observer.  In the presence of a cosmological constant $\Lambda$ the number of distinguishable universes is bounded by $e^{|\Lambda|^{-3/4}}$. In the context of the string theory landscape, the overall number of different universes is expected to be exponentially greater than the total number of vacua in the landscape. We discuss the possibility that the strongest constraint on the number of distinguishable universes may be related not to the properties of the multiverse but to the properties of  observers.

\end{abstract}
\maketitle

\section{Introduction}\label{intro}

With the invention of inflationary cosmology, the notion of a uniform universe was gradually replaced by the notion of a multiverse consisting of many locally uniform exponentially large parts  \cite{linde1982,linde1983}. Each of these parts locally looks like a uniform nearly-Friedmann universe. A collection of all of these universes represents an eternally growing fractal consisting of many such ``universes'' with different properties   \cite{Vilenkin:1983xq,Linde:1986fd,Linde:1987aa}. This scenario recently became quite popular when a mechanism to stabilize string theory vacua was found \cite{Kachru:2003aw}, and string theorists realized \cite{Douglas}, in agreement with earlier expectations \cite{Lerche:1986cx,Bousso:2000xa}, that the total number of different stringy vacua can be extremely large. The popular estimate of the number of different vacua is $\sim 10^{500}$, but the true number may be much smaller or much greater than that  \cite{Douglas}. Because of the transitions from one vacuum state to another, the inflationary multiverse becomes divided into an exponentially large number of different exponentially large ``universes'' with different laws of low-energy physics operating in each of them. This picture, which is now known as the string theory landscape \cite{Susskind:2003kw}, was envisaged in the very first paper on eternal chaotic inflation \cite{Linde:1986fd}.

But the string theory landscape does not fully describe all of the options which may exist in an  inflationary multiverse. The properties of our world are determined not only by the properties of the vacuum state; after all, we do not live in a vacuum. We live in the world  with a certain matter content and a particular large-scale structure. Even if we concentrate on a single vacuum state, i.e. on the same position in the string theory landscape, the large-scale structure (e.g. the spatial distribution of galaxies) and the matter content in each of the locally Friedmann parts of the cosmic fractal may be quite different. One may wonder how many  different locally Friedmann universes one may encounter in any particular part of the landscape.  Yet another question is how many locally distinguishable {\it classical} geometries of the universe one may encounter inside our cosmological horizon. 

Certain aspects of the multiplicity of the universe in the context of eternal inflation were previously discussed in Ref. \cite{Garriga:2001ch}. The authors were mainly interested in the number of possible outcomes which may appear inside a horizon-size classically uniform part of a post-recombination universe because of quantum or thermal fluctuations. Our goals are somewhat different: we will evaluate the number of inhomogeneous but locally Friedmann parts of the multiverse, which are different at a classical level. In a certain sense, which is going to be clear from the subsequent discussion, we will calculate the total number of classical histories for the geometry  of the universe. More precisely, we calculate the total number of different coarse grained classical post-inflationary initial conditions, which determine the subsequent evolution of the universe.

Our work is a part of the general effort towards finding the probability  to live in a universe with some particular set of properties. This  requires finding the probability measure in the multiverse, see e.g. \cite{Linde93,Linde94,Bousso,GarrigaVilenkin,GSPVW,Linde07,LVW,Page,Vanchurin,VanchurinVilenkin,VVW,Winitzki,GarciaBellido}. However, before embarking on that grand quest, it may be useful to solve a simpler problem:   to classify and count all possible universes. This is the main goal of our paper.

The paper is organized as follows. In Sec. \ref{s2} we estimate the entropy of  cosmological perturbations generated during slow-roll inflation and in Sec. \ref{eternal} we calculate the total number of distinguishable universes produced by eternal inflation. The number of universes in the presence of cosmological constant is evaluated in Secs. \ref{cc}  and in the context of the string theory landscape in Sec. \ref{land}.  In Sec. \ref{matter} we will briefly discuss entropy of normal matter, even though this entropy is not directly related to the number of possible classical geometries of the universe. In Sec. \ref{worlds} we argue that only a small fraction of all universes can even in principle be distinguished by a local observer. The main results are summarized in the Sec. \ref{conclusion}.

\section{Counting the universes}\label{s2}

When one starts thinking about the total number of various possibilities which exist in the multiverse, the first idea that comes to mind  is to take into account all quantum fluctuations, or real particles in a state of  thermodynamical equilibrium, which are present even in a totally homogeneous universe \cite{Garriga:2001ch}. For example, one could argue that the total number of all possible quantum configurations inside an event horizon of size $H^{{-1}} \sim \Lambda^{-1/2}$ in de Sitter space with the cosmological constant $\Lambda$ is given by $e^{S_{dS}}$, where  $S_{dS} = 24\pi^{2} \Lambda^{-1}$ is the Gibbons-Hawking dS entropy. Those who would like to deal with something more tangible than quantum fluctuations, could suggest to calculate the quantity $e^{S_{\rm matter}}$, describing the entropy of all particles in a given part of the universe. However, in this paper we are not going to count  short living  fluctuations, or Boltzmann brains, which may emerge because of exponentially improbable quantum (or thermal) fluctuations even in an absolutely uniform space without planets and galaxies. Thus we should turn our attention to something else.

According to inflationary cosmology, the large-scale structure of the universe, which is necessary for our existence, is a result of quantum effects which occurred at the stage of slow-roll inflation \cite{Mukh}. This is a distinguishing feature of the slow-roll inflation, as compared to the false vacuum inflation in different metastable dS vacua of the landscape.  

During  inflation with the Hubble constant $H_I$, quantum fluctuations of all scalar fields with masses $m <H_I$ are generated. These perturbations produced during a typical time $H_I^{{-1}}$ have a typical amplitude $\delta \phi  = \pm {H_I\over 2\pi}$ and a wavelength $O(H^{{-1}})$. When they are stretched to an exponentially large scale by inflation, they stop oscillating (freeze) and start looking as a nearly homogeneous {\it classical} scalar field. With each new e-fold of inflation, new perturbations are generated on top of the previously generated ones. This is the standard mechanism of production of perturbations of metric responsible for formation of the large scale structure of the universe. In this manner, quantum fluctuations during inflation prepare different {\it classical} initial conditions for the subsequent evolution of different parts of the universe.

Note that this process occurs independently in each part of the universe of size $H_I^{-1}$. Classical scalar fields produced by freezing and stretching of inflationary quantum fluctuations determine classical initial conditions for all physical processes in the post-inflationary universe, on an exponentially large scale. The properties  of quantum jumps determine the properties of the universe on a scale corresponding to the size of the initial Hubble size domain stretched by the subsequent cosmological evolution. 

If the inflaton field driving inflation jumps in the direction opposite to its classical motion (i.e. uphill), this produces a slightly overdense region on the corresponding scale; if the field jumps downhill, it produces an underdense region. In other words, geometric properties of our world are determined by the chain reaction of quantum jumps during inflation. 

Once inflation is over, it leaves its remnants in form of the large scale classical perturbations of metric and of various light scalar fields. If one plugs in all of these perturbations in a powerful computer and evolve them in accordance with the laws of physics operating in our part of the landscape, one should be able, at least in principle, to find all details of the subsequent galaxy formation, star formation, post-inflationary entropy production, etc. 

In other words, one can argue that {\it the evolution and properties of all important macroscopic features of the universe can be traced back to the two main ingredients:

1) The properties of our vacuum state, represented by one of the many vacua in the landscape.

2) The properties of the slow-roll inflation and of the large scale perturbations of metric and physical fields produced at that stage.}

The main idea of our paper is that once one finds all possible combinations of these two ingredients, one can determine all classical histories of the universe, and, consequently, all possible macroscopic features of our universe that are required for our existence.

There are some obvious exceptions from this conjecture. For example, if the slow roll inflation is very short, we may be affected by bubble collisions or other physical processes preceding  the stage of the slow roll inflation. In what follows, we will assume that the slow roll inflation is sufficiently long to allow us to ignore these processes, but in general on should take them into account. Also, the slow roll inflation is not the only process which generates classical fields in cosmology; another important example is preheating after inflation \cite{KLS}. The difference between these two processes is that  the growth of the occupation numbers of quantum fluctuations during the  post-inflationary preheating  has a power-law dependence on masses and coupling constants, whereas the growth of occupation numbers of particles during inflation is exponentially large. That is why we will concentrate on this process in our paper.

Now we will make an estimate of the total number of different classical geometries which may be produced by the slow-roll inflation. In our estimates we will make an important simplifying assumption. We will assume that the field make a single jump each time $H_I^{{-1}}$, and the magnitude  of the jump is $ \pm {H_I\over 2 \pi}$. In other words, we consider coarse-grained histories, ignoring, e.g., the   possibility that the field may, with an exponentially small probability, jump up or down by much more than $\pm {H_I\over 2 \pi}$.\footnote{This is a delicate issue since such trajectories may be important with some of the probability measures \cite{Linde94}. However, such measures suffer from the youngness paradox, so we will ignore this issue in our paper.} 

It is rather straightforward to calculate the total number of different coarse-grained geometries produced by this mechanism. Consider an inflationary domain of initial size $H_I^{{-1}}$ after it experienced $N$ e-folds of inflation. After the first e-fold the domain has grown $e$ times, and it contains now $e^{3}$ domains of size $H_I^{-1}$ in each of which the field could independently jump either by $+ {H_I\over 2 \pi}$ or by $- {H_I\over 2 \pi}$. The total number of different coarse-grained configurations of the field in this domain becomes $2^{e^{3}}$. Note that our estimate was very rough. We assumed that the field could experience only two possible jumps $\pm {H_I\over 2 \pi}$ and nothing in between (coarse-graining). Therefore our estimate is valid only up to a factor $ O(1)$ in the exponent, so one can write the final result as
$e^{C\,e^{3}}$, with  $C = O(1)$.

During the next time interval  $H_I^{{-1}}$  each of the $e^{3}$ domains of size $H_I^{-1}$ experience a similar set of jumps. They change the value of the scalar field on the scale $H_I^{-1}$, but do not change the results of the previous jumps on the scale $e H_I^{-1}$. The total number of different field configurations becomes $e^{C(e^{3}+e^{6})}$. Obviously, the total number of different configurations after $N$ e-folds of inflation becomes
\bea
{\cal N} &\sim& \exp \left(C \sum_{1}^{N} e^{3N}\right)\nonumber \\ &=& \exp \left({c e^{3N}}\right) \ ,
\eea
where $c$ is another constant $O(1)$. In what follows, we will write the final result in a simplified way,
\be
{\cal N} \sim e^{e^{3N}} \ ,
\label{eq:number}
\ee
keeping in mind the uncertainty in the coefficient in the exponent. Of course, for very large $N$ not all of these different universes will be seen by any particular observer; see a discussion of this issue in Section \ref{cc}. The estimate given above describes the total number of {\it all} possible universes which can be seen by {\it all} possible observers.

Note that the derivation of this result has a transparent physical interpretation: Each of the $e^{3N}$ independent inflationary domains has its own degree of freedom (the inflaton field inside it jumps either up or down).  This suggests that, up to a numerical factor, this system of long-wavelength perturbations produced during the slow-roll inflation has entropy $S_{{\rm infl}} \sim e^{3N}$. As we will see this entropy could be associated with the entropy of cosmological perturbations derived in Refs. \cite{BMP1,BMP2,Kiefer:1999sj,Podolsky:2005bw,Kiefer:2008ku,Campo:2008ij}. We should note that this entropy is totally different from the standard de Sitter entropy and from the entropy of normal matter.

To get a numerical estimate, suppose that our part of the universe was produced as a result of 60 e-folds of the slow-roll inflation of an inflationary domain of size $H_I^{-1}$.\footnote{In Sec. \ref{cc} we will show explicitly that the number of observable e-folds must be bounded by a logarithm of the cosmological constant (e.g. $N\lesssim 70$ in our universe).} This process may create 
\be
{\cal N} \sim e^{e^{180}} \sim 10^{10^{77}}
\ee
universes with different geometrical properties.  This number is incomparably greater than $10^{500}$. If the initial size of the universe is greater than $H_I^{{-1}}$, the total number of different universes is even much greater.

The diversity of various outcomes of inflationary evolution becomes even greater if there are more than one scalar field with the mass smaller than $H_I$. In such theories not only the local geometry but 
even the matter content of the universe in any given vacuum may also depend on inflationary quantum fluctuations. 
For example, the baryon/photon ratio $n_{B}/n_{\gamma}$ in the Affleck-Dine baryogenesis scenario \cite{Affleck:1984fy} depends on perturbations of the scalar field responsible for CP violation, and therefore it may take different values in different parts of an inflationary universe \cite{Linde:1985gh}. The ratio of dark matter to baryons $\rho_{DM}/\rho_{B}$ in axion cosmology is determined by long wavelength inflationary perturbations of the axion field, which takes different values in different parts of the multiverse \cite{Linde:1987bx,Tegmark:2005dy}.  
In the curvaton theory \cite{LM,Enqvist:2001zp,Lyth:2001nq,Moroi:2001ct}, the amplitude of perturbations of metric is different in different parts of the multiverse \cite{LM,LM2}. We will return to these possibilities later on.

In the remainder of this section we give an alternative interpretation of our results by following the analysis of Ref. \cite{BMP1}. The authors showed that both types of cosmological perturbations (gravitational waves and density perturbations) can be described by a stochastic scalar field $\phi$, whose entropy in the limit of large occupation numbers is given by
\begin{equation}
S\approx V\int d^{3}k\log \left(n_{\vec{k}}\right),\label{eq:entropy_inflation}
\end{equation}
where $n_{\vec{k}} $ is the number of particles. The spectrum of gravitational waves $\delta_h$ and density perturbations $\delta_\Phi$ is usually defined through corresponding two point correlation functions which could also be expressed through the average number of particles $\langle n_{\vec k} \rangle$. Therefore, it is a straightforward exercise to estimate the number of particles from a given spectrum of cosmological perturbations (see Refs. \cite{BMP2} for details). 

From (\ref{eq:entropy_inflation}) we can approximate the entropy of gravitational radiation contained inside volume $V=H_I^{-3} e^{3N}$ of the reheating surface
\begin{equation}
S_{gw} \sim H_I^{-3} e^{3N} \int_{H_I e^{-N}}^{H_I}k^2 dk \log\left(\frac{\delta_h a}{k}\right) \approx  e^{3 N} ,\label{eq:entropy_gw}
\end{equation}
where $H_{I}$ is the Hubble scale during inflation and $N$ is the number of e-folds of slow-roll inflation. The integral in (\ref{eq:entropy_gw}) is dominated by the high frequency modes $k\sim H_I$ indicating that most of the entropy is generated when a given mode crosses horizon, 
\be
\frac{S}{V} \sim H_I^{3}.
\label{eq:entropy_density}
\ee
Similarly, the entropy density of adiabatic perturbations is given by (\ref{eq:entropy_density}) up to a logarithmic correction \cite{BMP2}. In the limit of large occupation numbers the overall entropy in linear perturbations can be expressed as
\begin{equation}
S_{\rm pert} = c\ e^{3 N} ,\label{eq:total_entropy}
\end{equation}
where $c$ is some constant of order unity. It is now convenient to define the total number of universes as \begin{align}
{\cal N} & \equiv e^{S_{\rm pert}}  =  e^{c\, e^{3N}}, \label{eq:histories_defined}\end{align} 
which agrees qualitatively with our previous estimate (\ref{eq:number}). 

We should note that an accurate definition of the entropy of perturbations of metric requires a more detailed discussion; we refer the readers to the original literature on this subject, see e.g. \cite{BMP1,BMP2,Kiefer:1999sj,Podolsky:2005bw,Kiefer:2008ku,Campo:2008ij} and references therein. 
For the purposes of our paper, we will use the concept of entropy of perturbations of metric as a shortcut interpretation of our original estimate (\ref{eq:number}): $S_{\rm pert} \sim  \log{\cal N} \sim e^{3N}$. Another important comment here is that the main contribution to this number is given by the perturbations produced at the very end of inflation. These perturbations are only marginally ``classical.'' Therefore in order to use our estimates in a reliable way one should make a step back from the very end of inflation. This will somewhat reduce the extremely large numbers that we are going to discuss shortly. However, one may expect our estimates to be qualitatively correct in the large $N$ limit.

\section{Number of universes produced by eternal inflation}\label{eternal}

If  quantum jumps of the field $\phi$  dominate its classical rolling during a typical time  $H_I^{{-1}}$, then each domain of a size $H_I^{{-1}}$ will eternally split into many new domains, in some of which  the field will over and again jump against the classical rolling of the scalar field, forever re-starting the slow-roll process in different $H_I^{-1}$-sized domains. This leads to  eternal inflation~\cite{Vilenkin:1983xq,Linde:1986fd}. 
It occurs for fields satisfying the following generic condition \cite{Linde:1986fd}:
\bea
\langle\delta\phi\rangle_{\rm quant} & \gtrsim & \delta\phi_{\rm class}(\Delta t=H_I^{-1})=-\,\frac{V'}{3H_I^2}(\phi)\nn\\ &\Rightarrow&\; V^{3} \gtrsim 12 \pi^{2}(V')^{2} \ .
\label{eq:condition}
\eea

At the first glance, one could expect that the total number of different locally Friedmann universes produced by eternal inflation must be infinite since in this regime the number of e-foldings $N$ in (\ref{eq:number}) becomes indefinitely large. However, this is not the case. Indeed, quantum fluctuations which occur in the regime of eternal inflation produce perturbations of metric which are greater than $O(1)$ at the end of inflation \cite{book}. One can see it directly by comparing the condition required for eternal inflation (\ref{eq:condition}) with the amplitude of post-inflationary perturbations of metric, which are of the order ${V^{3/2} \over V'}$. Thus, all perturbations above the boundary of eternal inflation produce the universes which do not look like locally Friedmann universes, even approximately. That is why in order to find all nearly Friedmann universes produced by inflation it is sufficient to study the cosmological evolution of those parts of the universe where the condition (\ref{eq:condition}) is not satisfied and eternal inflation is over.

We will denote the boundary value of the field at which the  condition of slow roll eternal inflation  is satisfied as $\phi_{*}$. To calculate the total number of e-foldings after the end of eternal inflation in any particular part of the universe, we should take $\phi\sim\phi_*$ as the initial condition for the phase of slow-roll inflation, which  leads to a finite amount of slow-roll inflation \cite{Linde:2007jn}.

In slow-roll inflation the Hubble constant is given by $\sqrt{V/3}$, in Planck units, and  
\be  \label{Friedmann}
3H_I\dot \phi = - V' 
\ee
Using expression for de~Sitter entropy 
\be S = 24\pi^{2} V^{-1} = 8\pi^{2} H_I^{-2}  
\ee
and the relation $\dot N = H_I$ for the number of e-foldings $N$,
one can easily find that
\be
\frac{dS}{dN} = \frac{8 \pi^2 \dot \phi^2  }{H_I^4} 
\sim \bigg(\frac{\delta \rho}{\rho} \bigg)^{-2}
\ee
By integrating this equation, taking into account that $\frac{\delta \rho}{\rho} < 1$ and assuming that dS entropy at the end of the slow rolling is   larger than at the beginning, one can get a bound on the total number of $e$-foldings,
\be \label{Ntotbound}
N_{\rm tot} \lesssim  S_{\rm end} \ ,
\ee
where $S_{\rm end}$ is the Gibbons-Hawking de~Sitter entropy at the end of  slow roll inflation \cite{ArkaniHamed:2007ky}.

This is an interesting theoretical bound, but it is  not particularly informative in practical applications. Consider, for example, a simple model of the type of new inflation, or inflation near an inflection point,  with  potential
\be
V=V_0\left(1-\frac{\lambda_p}{p}\phi^p\right) .\label{Vsimp}
\ee
Note that here we absorbed $V_{0}$ in the definition of $\lambda_{p}$. To distinguish this case from the simplest versions of chaotic inflation scenario involving large fields $\phi> 1$, we will assume that $\lambda_{p} \gg 1$. In this regime, inflation begins at $\phi \approx 0$ and ends at $\phi\ll 1$.    In this situation the number of e-folds after eternal inflation is given by \cite{Linde:2007jn}
\be 
N_{\rm tot} \sim  \frac{(12\pi^{2})^{\frac{p-2}{2p-2}}}{p-2}\, \lambda_p^{-\frac{1}{p-1}} V_0^{-\frac{p-2}{2p-2}}\quad.\label{NtotAnalytp}
\ee
Consider for example the theory of the type of new inflation, with $V=V_0\left(1-\frac{\lambda_4}{4}\phi^4\right)$.
In this case one has
\be 
N_{\rm tot} \sim  \left( \lambda_4V_0\right)^{-\frac{1}{3}}\quad.\label{NtotAnalyt4}
\ee
One can show that for $\lambda_{4}> 1 $ and $V \ll 1$ the bound $N_{\rm tot} \lesssim  S_{\rm end}$ is satisfied in this scenario, but $S_{\rm end} \sim  V^{{-1}}$ is very much different from the actual number of e-foldings after the end of eternal inflation.

The situation is especially interesting and instructive in simplest models of chaotic inflation with 
\be
V= \frac{\lambda_n}{n}\phi^n  .\label{Vsimpchaot}
\ee
In this case,  the total number of e-folds since the end of eternal inflation can be estimated by 
\be
N_{\rm tot} \sim {2\phi_{*}^{2}/n} \sim 2 \left({12\pi^{2}\over \lambda_{n}}\right)^{2\over n+2} n^{1-n\over n+2}.\label{Nchaot}
\ee

In fact, one can easily check that in this class of theories
\be
N_{\rm tot} \sim C_{n}\, {S_{\rm e}},\label{Nchaot2}
\ee
where  $S_{\rm e}$ is the de~Sitter entropy at the boundary of eternal inflation and
\be
C_{n} = {n^{n-1\over n+2}\over 4}  = O(1)
\ee
for the simplest chaotic inflation models with $n = O(1)$.

To give a particular numerical estimate, in the theory $m^{2}\phi^{2}/2$, 
\be
N_{\rm tot} \sim  c\ m^{-1} \label{eq:total_efolds} .
\ee
where $c = 2^{5/3} 3^{1/2}\pi = O(20)$.
In realistic models one may expect $m \sim 3\times 10^{{-6}}$, and therefore
\be
N_{\rm tot} \sim 10^{7} \ .
\ee
Meanwhile the bound (\ref{Ntotbound}) in this case would be $N_{\rm tot} \lesssim m^{{-2}} \sim 10^{{11}}$, which is much weaker and less informative than the actual result $N_{\rm tot} \sim m^{{-1}}$ which we just obtained.

The total number of different types of universes produced in chaotic inflation with $V= m^{2}\phi^{2}/2$, $m \sim 3\times10^{-6}$, can be estimated by
\be
{\cal N} \sim e^{e^{3N_{\rm tot}}} \sim  e^{e^{3c/m}} \sim 10^{10^{10^{7}}}
\ee
This number may change significantly if we use a different definition of the boundary of the eternal inflation \cite{Winitzki2}, but with any definition, this number is VERY large. It is exponentially greater than the total number of  string theory vacua. This number may become even much greater if we take into account that the parameters of  inflationary models may take different values in different vacua in the landscape.

\section{Number of universes in the presence of the cosmological constant}\label{cc}

Not all of the universes produced since the end of eternal inflation can be distinguished by observers populating the observable part of the universe. 
%{\bf ***In this section we will limit ourselves to the discussion of the universes with $\Omega = O(1)$. Some comments concerning the case of strongly open universes with $\Omega \ll 1$ will be contained in Sect. \ref{land}.***}

During the post-inflationary expansion of the universe, each domain of initial size $H_{I}^{{-1}}$ grows as $H_{I}^{{-1}} a(t)$, where $a(t)$ is the scale factor, which is normalized to 1 at the end of inflation. At this stage the total size of the observable part of the universe grows approximately as $t$, so the total number of independent domains of initial size $H_{I}^{{-1}}$ accessible to observations (i.e.  
the total entropy of observable cosmological perturbations) grows as
\be
S_{\rm pert}(t) \sim \left(\frac{t H_I }{a(t)}\right)^3 \ .
\label{eq:entropy_approximate}
\ee
This regime continues only until the moment when the energy density of all matter becomes smaller than the absolute value of the cosmological constant $\Lambda$. 

For $\Lambda > 0$, starting from the time $t \sim \Lambda^{{-1/2}}$ the universe starts expanding exponentially, and we no longer see new parts of the universe, which leads to a cutoff in the observable information stored in the cosmological perturbations. Meanwhile for $\Lambda < 0$ the universe typically collapses within the time $t \sim |\Lambda|^{{-1/2}}$. Thus in both cases in order to estimate the total entropy of observable cosmological perturbations it is sufficient to limit ourselves to what one can observe within the cosmological time $t \sim |\Lambda|^{{-1/2}}$.

At $t \sim |\Lambda|^{{-1/2}}$, the energy density of gravitational waves, which contribute only a fraction to the overall matter density, must be strictly smaller than the absolute value of the cosmological constant,
\be
\rho_{gw} = H_I^{4}a^{-4}<\left|\Lambda\right| \ . \label{eq:gw_density}
\ee

The above bound can be saturated only if the energy density of gravitational waves dominate the energy density of all other types of matter at  the epoch when this energy density decreases and approaches the value comparable to $\left|\Lambda\right|$. By combining (\ref{eq:entropy_approximate}), (\ref{eq:gw_density}) and using $H=\sqrt{\left|\Lambda\right|/3}$ we find
\be
S_{\rm pert} \lesssim   \left|\Lambda\right|^{-3/4}. \label{eq:entropy_bound}
\ee

One should note that our estimates are valid for the universes which have geometry not too different from the geometry of a flat universe, $\Omega = O(1)$. If one considers open universes with $\Omega \ll 1$, which are very different from the universe where we live now, the total entropy  inside the observable part of the universe may be much greater than $|\Lambda|^{-3/4}$, approaching $\Lambda^{-1}$ for $\Lambda > 0$ and $\Lambda^{-2}$ for $\Lambda > 0$ \cite{Bousso:2010pm}. We will not study this regime in our paper.

It follows from (\ref{eq:total_entropy}) and  (\ref{eq:entropy_bound}) that the maximum number of observable e-folds is bounded by
\be
N_{\rm max} \sim  - \frac{\log(\left|\Lambda\right|)}{4}.
\label{eq:efolds_max}
\ee
This number is typically much smaller than the total number of e-folds estimated in (\ref{eq:total_efolds}). For example, the maximum number of observable e-folds in our vacuum with $\Lambda \sim 10^{-120}$ is about 70, which is pretty close to what is actually observed.

At first, it could seem that the bound (\ref{eq:entropy_bound}) can always be saturated regardless of the scale of inflation. However, usually this is not the case. Suppose inflation ends at $a = 1$ and after an instant stage of reheating the universe becomes dominated by matter with $p_{w} = w \rho_{w}$. Then, at the time when the density of matter becomes comparable to the value of the cosmological constant  one has
\be
\rho_{w} = H^2_I a^{-3(1+w)}\approx |\Lambda| \approx t^{-2} \ . \label{w}
\ee
From (\ref{eq:entropy_approximate}), one finds that the maximal value of the observable entropy is
\be
S_{\rm pert} \approx  H_I^{\frac{1+3w}{1+w}} \left|\Lambda\right|^{-\frac{1+3w}{2+2w}}. \label{w2}
\ee

In order to analyze a particular semi-realistic example, consider the universe dominated by relativistic matter soon after the end of inflation ($w = 1/3$). In this case
\be
S_{\rm pert} \sim   H_I^{\frac{3}{2}} \left|\Lambda\right|^{-{3}/{4}}.
\label{eq:entropy_pert}
\ee
In this regime the bound (\ref{eq:entropy_bound}) is saturated if inflation ends at the Planck density, $H^{2}_{I} = O(1)$.\footnote{Looking at Eq. (\ref{w2}), one could expect that, for example, for the stiff equation of state $w = 1$ one could have entropy $O(\Lambda^{-1})$, which is much greater than the bound (\ref {eq:entropy_bound}). However, one can show that in this case the energy of gravitational waves eventually begins to dominate and the bound (\ref {eq:entropy_bound}) holds, as it should.}  However, in realistic models of inflation with $H^{2}_{I} \lesssim 10^{{-9}}$ one finds $S_{\rm pert} \sim H_I^{\frac{3}{2}} \left|\Lambda\right|^{-{3}/{4}} \ll  \left|\Lambda\right|^{-{3}/{4}}$. In particular, in the simplest chaotic inflation model with $V= m^{2}\phi^{2}/2$, $m \sim 3\times10^{-6}$, and $\Lambda \sim 10^{-120}$ the maximal contribution to the entropy is given by the last stage of inflation where $H_{I}\sim m \sim  3\times10^{-6}$, so one finds (assuming instant reheating and $w = 1/3$):
\be
S_{\rm pert} \sim   5\times 10^{{-9}} \left|\Lambda\right|^{-{3}/{4}} \sim 10^{82}.
\label{eq:entropy_pert1}
\ee
which gives the total number of different universes
\be
{\cal N} \sim   10^{10^{82}}.
\label{eq:entropy_pert2}
\ee
Of course this is a very rough estimate. In particular, as we already mentioned, the largest contribution to this number is given by perturbations produced at the latest stages of inflation. Such perturbations do not have much time to inflate and their occupation numbers are not exponentially large, unless one makes a sufficiently large step back from the end of inflation, which effectively decreases the number of e-foldings contributing to our estimate. Moreover, one may argue that the information about the last few e-foldings of inflation may be erased by subsequent cosmological evolution. This may somewhat reduce the estimated power 82 in (\ref{eq:entropy_pert2}), but the total number of possible observable universes will remain extremely large.

Before we discuss a similar result in the context of the string theory landscape, we should note that one may be interested not in what could be potentially possible in the unlimited future, but in what is possible within some important range of time. If, for example, we are interested in the total number of options for the observable part of the universe with age $t \sim 10^{10}$ years, then the results will be essentially the same as in the models with  $\Lambda \sim 10^{-120}$.  

\section{Number of universes in the Landscape}\label{land}

Now let us estimate the number of distinct universes in the entire landscape. If we assume that the total number of vacua is $M$, then from (\ref{eq:entropy_bound}) the total number of distinct universes is given by a sum over all vacua
\begin{equation}
{\cal N}\approx \sum_{i=1}^{M}e^{\left|\Lambda_{i}\right|^{-\frac{3}{4}}}.\label{eq:all_histories}
\end{equation}
Here, in order to make a rough estimate,  we assumed that the upper bound (\ref{eq:entropy_bound}) can be saturated. Clearly, the largest contribution to the number of universes (\ref{eq:all_histories}) comes from the vacua with the smallest absolute value of the cosmological constant.

As we already mentioned, the popular estimate for $M$ is $10^{500}$, but in fact it can be much smaller or much greater than that. Assuming for simplicity that the vacua are flatly distributed near $\Lambda =0$, one may expect that the lowest nonvanishing value of $\Lambda $ is $|\Lambda_{\rm min}| \sim 1/M$. Then from our estimates it would follow that the maximal number of observable e-folds is $N_{\rm max}\sim 290 $ and the corresponding number of distinct universes is
\begin{equation}
{\cal N}\sim e^{\left|\Lambda_{\rm min}\right|^{-\frac{3}{4}}} \sim e^{M^{\frac{3}{4}}}   \sim10^{10^{375}}.\label{eq:all_histories2}
\end{equation}
As we see, the total number of the observable geometries of the universe  is expected to be exponentially greater than the total number $M$ of  string theory vacua in the landscape: ${\cal N}\sim e^{M^{\frac{3}{4}}}$.

But what if the minimal value of $\Lambda$ in the landscape is $\Lambda = 0$? This is a viable possibility. In fact, one of the vacua in string theory landscape, which corresponds to the decompactified 10D universe, does have $\Lambda = 0$. Does this mean that an observer in such vacua will see an infinite number of universes?

The answer is that for very small $\Lambda$ we would be able to see the universe on the scale corresponding to the maximal number of e-folding in the slow-roll regime, or on the scale corresponding to the boundary of self-reproduction. In the last case, the total number of different observable universes will be given by ${\cal N}$ estimated in Section \ref{eternal}.

The situation becomes a bit more subtle in an open universe if the slow-roll evolution is not eternal. Estimates given in the previous sections rely on the assumption of flatness $\Omega \sim O(1)$, meanwhile, as we already mentioned, for $\Omega \ll 1$, one can find an even much greater number of distinguishable universes, with $\Lambda^{-3/4}$ in (\ref{eq:all_histories}) replaced by $\Lambda^{-1}$  or  $\Lambda^{-2}$  for positive or negative values of $\Lambda$ respectively \cite{Bousso:2010pm}. Moreover, in the models with vanishing cosmological constant the observable area of the reheating surface is not bounded from above. An observer in such universe (sometimes called the census taker \cite{Susskind}) could in principle count an infinite number of distinguishable universes. However, the only known vacuum with exactly zero vacuum energy is the vacuum which appears when the universe decompactifies and become 10D. In general, supersymmetric 3D vacua may exist in string theory landscape \cite{Kallosh:2004yh}, but as of now we do not see any reason for the tremendous fine-tuning required for this to happen. As we already mentioned,  if the vacua are flatly distributed near $\Lambda =0$, one may expect that the lowest nonvanishing value of $\Lambda $ is $|\Lambda_{\rm min}| \sim 1/M$, which was the origin of our estimate  (\ref{eq:all_histories2}) of the number of different universes with $\Omega = O(1)$: \, ${\cal N}\sim e^{M^{\frac{3}{4}}}$.

In our estimates in the last two sections we made an assumption that local properties of our universe cannot be affected by fluctuations on the scale much greater than the present horizon. This assumption can be violated in the theories with more than one light scalar field. For example, quantum fluctuations of the axion field during inflation may produce perturbations with the wavelength many orders of magnitude greater than the size of the observable horizon. Inside our part of the universe, the sum of all such  perturbations can be interpreted as a homogeneous axion field. This field  determines the initial value of the axion field at the onset of the axion oscillations, and, as a result, it determines the ratio of 
dark matter to usual matter in our universe \cite{Linde:1987bx}. If we 
follow only the degrees of freedom inside our horizon, we may miss this fact, as well as the possibility to explain the present ratio of the dark matter to normal matter by anthropic considerations \cite{Linde:1987bx,Tegmark:2005dy,Freivogel:2008qc}. The same is true with respect to some other effects which we mentioned in section \ref{s2}, such as the possibility to give an anthropic explanation of the baryon asymmetry of the observable part of the universe in the Affleck-Dine scenario \cite{Affleck:1984fy,Linde:1985gh} and the possibility to explain the amplitude of perturbations of metric in the curvaton scenario \cite{LM,LM2}. 

One way to take into account this missing information during the counting of all possible universes is to apply the coarse-graining ideology. For example, during eternal inflation in the axion theory, the axion field becomes distributed all over the periodic phase space of its values, from $0$ to $2\pi f_{a}$, where $f_{a}$ is the radius of the axion potential. In terms of the coarse-grained histories, this dispersion may be represented as consisting of $4\pi^{2} f_{a}/H_I$ intervals of length ${H_I\over 2\pi}$. If this interval were in the range of wavelengths within our horizon, it would contribute an exponentially large factor to the number of possible universes. Inside the horizon we do not have any information about the exact history of perturbations with superhorizon wavelength, but we still have a factor $4\pi^{2} f_{a}/H_I$ describing a family of coarse-grained possibilities for the locally observable properties of the universe filled by a classical oscillating axion field. This extra factor is not exponentially large, but if one ignores it, one could miss one of the most interesting anthropic predictions of the theory of inflationary multiverse.

\section{Entropy of matter}\label{matter}

So far we were mainly interested in the distinct classical geometries produced by inflationary perturbations of metric.  These perturbations provide the set of classical initial conditions for the subsequent evolution of the universe. In our study we concentrated on perturbations with an amplitude smaller than $O(1)$, which produce locally Friedmann parts of the universe. For completeness, we will briefly discuss here the entropy of the usual matter, and also the entropy which can be produced when perturbations of metric become large, which leads to black hole production and their subsequent evaporation. Various issues related to the contents of this section have been discussed in many classical papers on cosmology, as well as in more recent works including Refs. \cite{Bousso:2007kq,EL}. 

First of all, let us estimate the maximal amount of entropy of normal matter which can be accessible to an observer in a universe with $\Omega = O(1)$ and a cosmological constant $\Lambda$. The most important constraint here comes from the fact that for $\Lambda > 0$, particles leave the observable part of dS space within the time $t \sim \Lambda^{{-1/2}}$, whereas for $\Lambda < 0$ the universe typically collapses within the time $t \sim |\Lambda|^{{-1/2}}$, so in both cases in order to estimate the total entropy produced after reheating that one can observe, it is sufficient to limit ourselves to what one can observe within the cosmological time $t \sim |\Lambda|^{{-1/2}}$.

The total entropy of the universe will be maximized if at the time when the energy density becomes comparable with the cosmological constant, all particles are ultrarelativistic. Assuming, for simplicity, that the number of types of massless particles is $O(1)$, one finds $T^{4} \sim |\Lambda|$ and the total entropy within a sphere of radius $|\Lambda|^{{-1/2}}$ is 
$S_{\rm matter}\sim  |\Lambda|^{{-3/4}}$. 
If the energy density at $t \sim |\Lambda|^{{-1/2}}$ is not dominated by ultrarelativistic particles, the total entropy of observable matter will be only smaller than $|\Lambda|^{{-3/4}}$, so one has a bound
\be
S_{\rm matter}\lesssim  |\Lambda|^{{-3/4}} \ ,
\ee
The same result can be obtained using the results of Ref. \cite{tHooft}, see  \cite{Boussotalk}. This bound is very similar to the upper bound on the entropy of inflationary perturbations (\ref{eq:entropy_bound}). However, one can obtain a more precise and stringent bound for the entropy of inflationary perturbations, which depends not only on $\Lambda$ but also on the Hubble constant at the end of inflation: $S_{\rm pert} \lesssim   H_I^{\frac{3}{2}} \left|\Lambda\right|^{-{3}/{4}}$ (\ref{eq:entropy_pert}).

In our universe, the upper bound $S_{\rm matter}\lesssim  |\Lambda|^{{-3/4}}$ would yield the total entropy of particles $S_{\rm matter} \sim  10^{{90}}$. However, in reality the energy density of photons is several orders of magnitude smaller than the energy density of baryons, which is about 5 times smaller that the energy density of dark matter. Therefore the total entropy of particles populating the observable part of our universe is several orders of magnitude smaller than its upper bound $O(|\Lambda|^{{-3/4}})$:\, $S_{\rm matter} \sim 10^{88}$. Thus, the main reason why the upper bound is not exactly saturated lies in the fact that ultrarelativistic matter (photons, gravitons etc.) contribute only a small fraction to the total energy density of the universe, as compared to baryonic matter and dark matter, at the moment when this density drops down to $|\Lambda|$. It is interesting that the ratio of the energy density of photons to energy density of nonrelativistic matter may have an anthropic origin \cite{Linde:1985gh,Linde:1985gh,Linde:1987bx,Tegmark:2005dy,Freivogel:2008qc}. Thus,  anthropic considerations may explain the reason why the upper bound on the entropy of particles is almost saturated in our universe.

As we already mentioned before, the total entropy of inflationary perturbations in the observable part of our universe is expected to be  further from saturating the upper bound $|\Lambda|^{{-3/4}}$, see Eq. (\ref{eq:entropy_pert1}).

One should note, that once the perturbations of metric grow and black holes form and evaporate, the total entropy inside the observable part of the universe may considerably increase. This is what happens in our universe, where
the entropy of black holes in the centers of galaxies can be greater than $10^{104}$ \cite{EL}. In particular, the entropy of a single black hole at the center of our own galaxy can be greater than the entropy of all particles in the observable part of the universe!

In the long run, most of the neighboring galaxies will move further and further away from our galaxy due to cosmic acceleration. If our galaxy (together with Andromeda) will eventually collapse into a single gigantic black hole,  its entropy will approach
\be
S_{\rm MilkyHole} \sim 10^{100} \ .
\ee
Of course, the entropy will be much smaller if some parts of matter in our galaxy form many smaller black holes which will evaporate earlier. Moreover,  it would take a very long time for the Milky Hole to form and an even longer time for us to observe its entropy in the form of Hawking radiation. It is interesting, nevertheless, that the total entropy produced by all localized objects in the observable part of our universe can be totally dominated by the entropy produced by the black hole evaporation.

If instead of considering our part of the universe we will consider all regimes that are possible in the landscape, one may envisage the possibility that the total entropy of a cosmological black hole may approach the dS entropy $O(\Lambda^{{-1}})$. This may happen, for example, if the amplitude of density perturbations on length scale $\sim |\Lambda|^{{-1/2}}$ can be $O(1)$; see a related discussion in \cite{Bousso:1996au,GarciaBellido:1996qt}.

\section{Counting worlds and minds}\label{worlds}

In our calculations of the total number of different universes in the previous sections, we were assuming that because the large scale fluctuations of the scalar field can be interpreted as classical fields, all of the different universes produced by eternal inflation have some kind of real, observer-independent existence. However, each time the meaning of these words was somewhat different.

When we were talking about all possible universes produced during eternal inflation, we counted everything that could be measured by all kinds of observers which may live everywhere in the multiverse. In other words, we counted all possible classical or semiclassical configurations, all possible histories, not only the ones associated with the observable events inside the cosmological horizon.

When we started talking about the universes inside the horizon, we paid attention to the fact that the total number of outcomes which can be registered by any particular observer at any moment of time is smaller than the total number of possibilities which could emerge in all parts of the universe. For example, an observer living inside a horizon-size patch of an exponentially expanding universe does not have access to other parts of the universe. Therefore some authors argue that anything that happens outside the horizon should not play any role in our counting of the universes and evaluation of probabilities. 

We do not want to discuss here validity of this argument. Instead of that, we would like to note that there are additional quantum mechanical limitations on what can be actually observed by any local observer.
For example, when one considers the Schrodinger cat experiment, this experiment has two definite outcomes: the cat can be either dead or alive. However, in accordance with the Copenhagen interpretation, these potentialities become realized only after one of these outcomes becomes registered by a classical observer. In the many-world (relative state) interpretation of quantum mechanics, we are talking about correlations between various observations made by an observer and the state of the rest of the universe. 

In everyday life, observers are big and very much classical, so their quantum nature can be safely ignored. However, the crucial ingredient of our procedure of counting the universes was an investigation of quantum effects on a wide range of scales from Planck length to supergalactic scales. Meanwhile each of us is $10^{26}$ times smaller than the cosmological horizon and $10^{35}$ times larger than the Planck scale. Thus one may wonder to which extent one can talk about a classical observer when discussing quantum effects on the scales much smaller or much greater than the size of an observer. Are there any constraints on the total number of {\it distinguishable} universes which are related to the quantum nature of an observer? 

This issue becomes manifest when one remembers that the essence
of the Wheeler-DeWitt equation, which is the Schr\"{o}dinger
equation for the wave function of the universe, is that this wave
function  {\it does not depend on time}, since the total
Hamiltonian of the universe, including the Hamiltonian of the
gravitational field, vanishes identically \cite{DeWitt:1967yk}. 

The resolution of this paradox  suggested by   Bryce
DeWitt \cite{DeWitt:1967yk} is rather instructive. The notion
of evolution is not applicable to the universe as a whole since
there is no external observer with respect to the universe, and
there is no external clock  that does not belong to the
universe. However, we do not actually ask why the universe {\it
{as a whole}} is evolving. We are just trying
to understand our own experimental data. Thus, a more precisely
formulated question is {\it why do   we see } the universe
evolving in time in a given way. In order to answer this question
one should first divide the universe into two main pieces. The first part consists of an
observer  with his clock and other measuring devices, with a combined mass $M$ and a total energy $M c^{2}$. The second part is the rest of the universe, with the total energy $-Mc^{2}$. Since the Hamiltonian (the energy) of the rest of the universe does not vanish, the wave function of the rest of the universe does depend on the state of the clock
of the observer, i.e. on his `time'. 

One of the implications of this result is that one can talk about the evolution of the universe only with respect to an observer. In the limit when the mass of the observer vanishes, the rest of the universe freezes in time. In this sense, the number of distinct observable histories of the universe is bounded from above by the total number of the histories that can be recorded by a given observer. And this number is finite.

Indeed, the total number ${\cal N}$ of all observable universes which could be recorded by a given observer is bounded from above by $e^{I}$, where $I$ is the maximal information that he/she can collect.
For any observer of mass $M$ and size $R$, this information cannot exceed the Bekenstein bound
\be I <S_{\rm Bek} = 2\pi MR.\ee
This bound implies that 
\be
{{\cal N}_{\rm observer}} < e^{S_{\rm Bek}} = e^{2\pi MR} \ .
\ee
For a typical observer with $M \sim 10^{2}$ kg and $R \sim 1$ m, one finds
\be
{{\cal N}_{\rm observer}} \lesssim e^{10^{45}} \ .
\ee
%Thus the number of possible quantum states of an observer of a human size is much smaller than the total number of the universes discussed in the previous sections.

Moreover, if we consider a typical human observer, the total amount of  information he can possibly absorb during his lifetime is expected to be of the order of $10^{16}$ bits or so \cite{DeSimone:2008if}. In other words, a typical human brain can have about 
\be
{{\cal N}_{\rm observer}} \sim
10^{10^{16}}\ee
different configurations, which means that a human observer may distinguish no more than $10^{10^{16}}$ different universes. This is a huge number, which is much greater than the standard estimate of the number of  dS vacua in the landscape $10^{500}$. However, this number is much smaller than the total number of possible geometries of the universe inside the cosmological horizon after 60 e-folds of inflation.

Thus we are discussing an additional constraint which previously did not attract much attention: The total number of possibilities accessible to any given observer is limited not only by the entropy of perturbations of metric produced by inflation and by the size of the cosmological horizon, but also by the number of degrees of freedom {\it of an observer}. This number   is tremendously large, so one can safely ignore this limitation in his/her everyday life. But when we study quantum cosmology, evaluate the total number of the universes and eventually apply these results to anthropic considerations, one may need to take this limitation into account. Potentially, it may become very important that when we analyze the probability of existence of a universe of a given type, we should be talking about a consistent pair: the universe and an observer who makes the rest of the universe ``alive'' and the wave function of the rest of the universe time-dependent. 

\section{Conclusion}\label{conclusion}

In this paper we made an attempt to find out how many different coarse-grained universes could be produced by inflation in each particular vacuum, and in the string theory landscape as a whole. The meaning of these words can be explained as follows. Slow-roll inflation produces long-wavelength perturbations of the metric, which become imprinted on the cosmological background and determine the large scale structure of the universe. Even though these perturbations are created from quantum fluctuations, they become essentially classical due to inflation. These perturbations provide different classical initial conditions in different parts of the universe. Our goal was to estimate the number of distinctly different classical geometries which may appear as a result of this effect. We found that the result is proportional to $e^{e^{3N}}$, where $N$ is the number of e-foldings of slow-roll inflation. This aspect allows one to look from a different perspective on the possible significance of slow-roll inflation, which helps to create the information content of the universe.

The estimate of the total number of distinct geometries produced by inflation depends on the method by which one can make this distinction. In the first part of this paper we concentrated on investigation of all possible locally Friedmann geometries which can be produced after the end of eternal inflation. Our goal was to understand how many different locally-Friedmann   (i.e. approximately homogeneous and isotropic) universes constitute the multiverse, which, as a whole, looks like a very inhomogeneous and anisotropic non-Friedmann eternally growing fractal. We found that the total number of such universes, in the simplest inflationary models, may exceed $10^{10^{10^{7}}}$. This humongous number is strongly model-dependent and may change when one uses different definitions of what is the boundary of eternal inflation.

Then we decided to limit ourselves to only those universes which can be distinguished from each other by a local observer in a universe with a given cosmological constant $\Lambda$. The resulting number appears to be  limited by $e^{\Lambda^{-3/4}}$. If this limit can be saturated, then the total number of locally distinguishable configurations in string theory landscape can be estimated by $e^{M^{3/4}}$, where $M$ is the total number of vacua in string theory. In other words, the total number of locally distinguishable geometries is expected to be exponentially greater than the total number of vacua in the landscape.

Finally, we checked how many of these geometries can be actually distinguished from each other by a local observer of given mass and size. Not surprisingly, since any local observer is smaller than the observable part of the universe, we have found that the strongest limit on the number of different locally distinguishable geometries is determined mostly by our own abilities to distinguish between different universes and to remember our results.

In this paper we did not attempt to draw deep philosophical conclusions based on our estimates, or apply them immediately to the search for the probability measure in the multiverse. Just as those who calculated the number of all possible vacua in the landscape, we concentrated  on finding some facts, leaving their interpretation for further investigation.  For example, it might be worthwhile to explore some simple measures which could emerge from our discussion. What would be the observational predictions if each of the universes have  the probability to be observed $P={1 \over {\cal N}}$? What if the probability is proportional to the observable entropy of inflationary perturbations $P\propto S$? Is it possible to apply our results to the stationary measure \cite{LVW, Linde:2007nm}? We are planing to return to these and other related issues in the future. 

\section*{Acknowledgments}

The authors are grateful to Jaume Garriga, Lev Kofman, Slava Mukhanov, Alex Vilenkin, Alexander Westphal and Sergei Winitzki for helpful discussions.
The work of A. L. was supported in part by NSF grant PHY-0244728, 
by the Alexander-von-Humboldt Foundation, and  by the FQXi grant RFP2-08-19. The work of V. V. was supported
in part by FQXi mini-grants MGB-07-018 and MGA-09-017.

\end{document}